\documentclass[preprint,number]{elsarticle}
\usepackage[left=2.54cm,top=2.54cm,right=2.54cm,nohead]{geometry}
\usepackage{natbib}
\usepackage{url}
\usepackage{setspace}
\usepackage[dvips]{epsfig}
\usepackage{color}
\usepackage{amsmath}
\usepackage{hyperref}

\usepackage{subfig}
\usepackage{booktabs}

\begin{document}

\begin{frontmatter}
\title{Free Surface Lattice Boltzmann with Enhanced Bubble Model}

\author[LSTM]{Daniela Anderl\corref{cor1}}
\ead{daniela.anderl@lstm.uni-erlangen.de}

\author[LSS]{Simon Bogner}

\author[LSTM]{Cornelia Rauh}

\author[LSS]{Ulrich R\"ude}

\author[LSTM]{Antonio Delgado}

\address[LSTM]{Lehrstuhl f\"ur Str\"omungsmechanik,
	Universit\"at Erlangen-N\"urnberg,
	Cauerstra{\ss}e 4,
	91058 Erlangen}

\address[LSS]{Lehrstuhl f\"ur Systemsimulation, 
	Universit\"at Erlangen-N\"urnberg,
	Cauerstra{\ss}e 11,
	91058 Erlangen}

 \cortext[cor1]{Corresponding author}

\begin{abstract}
  This paper presents an enhancement to the free surface lattice
Boltzmann method (FSLBM) for the simulation of bubbly flows including
rupture and breakup of bubbles. The FSLBM uses a volume of fluid
approach to reduce the problem of a liquid-gas two-phase flow to a
single-phase free surface simulation. In bubbly flows compression
effects leading to an increase or decrease of pressure in the
suspended bubbles cannot be neglected. Therefore, the free surface
simulation is augmented by a bubble model that supplies the missing
information by tracking the topological changes of the free surface in
the flow. The new model presented here is capable of handling the
effects of bubble breakup and coalesce without causing a significant
computational overhead. Thus, the enhanced bubble model extends the
applicability of the FSLBM to a new range of practically relevant
problems, like bubble formation and development in chemical reactors
or foaming processes.

\end{abstract}


\end{frontmatter}

\section{Introduction}

Splitting of bubbles occurs in many industrial applications. 
In food industry, foams are generated by gas inblow through a membrane which leads to a detachment, breakup and coalescence process of bubbles. 
To simulate such processes, the effect of bubbles splitting and coalescing in the flow needs to be treated correctly.
This paper suggests a three-dimensional \emph{free surface lattice Boltzmann method} (FSLBM) \cite{Ginzburg2003,KoernerEtAl} approach with an enhanced \emph{bubble model} capable of covering the breakup of a bubble.
The lattice Boltzmann method (LBM) \cite{Succi,Wolf-Gladrow,ChenDoolen1998,BenziEtAl} allows a straightforward parallelization and is hence capable to simulate huge scenarios like the development of foam at a membrane in principle. 
The FSLBM is based on a \emph{volume of fluids} approach \cite{HirtNichols,ScardiovelliZaleski} that treats the free surface as a free boundary in the fluid simulation. 
This approach can simplify the treatment of a liquid-gas two phase flow by assuming the second phase to be so light that it can be neglected.
Thus, the second phase does not account for the overall computational costs and the problem of high density ratios, which is a limiting factor in lattice Boltzmann multiphase approaches \cite{AidunClausen}, becomes obsolete. In recent years the FSLBM has been successfully applied to a number of problems like material foaming \cite{KoernerEtAl}, droplet motions \cite{XingEtAl2007}, rising bubbles \cite{DonathEtAl2010}, and offshore dynamics with grid refinement \cite{JanssenCAMWA2010,JanssenOffshore2010}.

Within free surface flow simulations there have already been approaches to treat bubbly flows with a large number of \emph{bubbles}, i.e. non-connected regions of gas, by the introduction of a \emph{bubble model} \cite{Caboussat2005,Donath2009}. 
While the dynamics within the bubbles is neglected, the gas pressure of a bubble due to compression is still taken into account by tracking the development of the volume of each separate bubble within the flow. 
The bubble model updates the pressure of each enclosed gas region according to the ideal gas law. Difficulty arises when the connectivity of the gas sub-domain changes, i.e., when interface advection leads to coalescence or breakup of bubbles. 
Previously, an implementation has demonstrated a bubble model including coalescence of bubbles in parallel FSLBM computations \cite{Donath2009}. The model of \cite{Cobussat2005} is capable of handling bubble breakup as well as coalescence, but needs an additional implicit solver for the bubble tracking.
In the following, we introduce an enhanced bubble model that also includes the coalescence and breakup of bubbles without causing additional computational costs. The capability of the new approach is demonstrated in a number of test cases, including the problem of bubble detachment from a pore. 


Sec. \ref{sec:LBM} and Sec. \ref{sec:FSLBM} introduce the used
numerical methods briefly.  In Sec. \ref{sec:bubbleModel} the
underlying bubble model and the algorithmic treatment of
bubble coalescence and breakup are explained. Implementation details for the bubble
model are given in Sec. \ref{sec:impl}. Also aspects of parallel
computation are discussed here. Finally Sec. \ref{sec:exp}
demonstrates the functionality of the proposed model. The two
application examples are the rupture of a gas thread into single
spherical bubbles due to capillary forces and the detachment of gas
bubbles from a circular orifice, as previously studied by
\cite{Gerlach2005,Gerlach2007}. We conclude in Sec. \ref{sec:conclusion}.

\section{Computational Model}
\label{sec:model}
\subsection{Lattice Boltzmann Method (LBM)}
\label{sec:LBM}

The LBM can be interpreted as a discretized Boltzmann equation \cite{Wolf-Gladrow,HeLuo1997} with a simplified collision term based on statistical mechanics
\begin{align}
	\frac{\partial f}{\partial t} + \boldsymbol{\xi} \cdot \nabla f = Q(f,f).
\end{align}
Here, $f(\boldsymbol{x},\boldsymbol{\xi},t)$, the probability density function, is the probability to meet a particle with velocity $\boldsymbol{\xi}$ at position $\boldsymbol{x}$ at time $t$ \cite{Haenel,Harris}.


For the collision term $Q(f,f) = \omega(f^{eq}-f)$ a model from \emph{Bhatnagar, Gross and Krooks} can be chosen to simplify the complex collision term with a relaxation towards equilibrium with the collision frequency $\omega$,
where $f^{eq}$ is the local equilibrium distribution, given by a Maxwell distribution.

To discretize the velocity space a certain number of degrees of freedom is necessary.
Here we use the three dimensional lattice model $D3Q19$, where $D3$ denotes three dimensions and $Q19$ the number of discrete velocities \cite{QianEtAl1992}.
The discretized \emph{particle distribution functions} (PDFs) are denoted in the following by
\begin{align}
	f(\boldsymbol{x},\boldsymbol{\xi},t) \rightarrow f(\boldsymbol{x},\boldsymbol{c}_\alpha,t)=f_\alpha(\boldsymbol{x},t).
\end{align}
For the $D3Q19$ model $\alpha$ ranges from $0$ to $18$.
The discrete lattice velocities are denoted by $\boldsymbol{c}_\alpha$.
To discretize the Maxwell distribution, a Taylor expansion for low $Ma$ numbers is used \cite{HeLuo1997} yielding
\begin{align}
 f^{eq}_\alpha = \rho w_\alpha \left[1 + \frac{\boldsymbol{c}_\alpha \cdot\boldsymbol{u}}{c_s^2} + \frac{(\boldsymbol{c}_\alpha\cdot \boldsymbol{u})^2}{2c_s^4} - \frac{\boldsymbol{u} \cdot \boldsymbol{u}}{2 c_s^2}\right]
\end{align}
where the $w_\alpha$ are weights according to the used lattice discretization.

The macroscopic quantities are calculated by taking the moments of the PDFs
\begin{align}
	\rho &= \sum_\alpha{f_\alpha}\\
	\rho \boldsymbol{u} &= \sum_\alpha{\boldsymbol{c}_\alpha f_\alpha} 
\end{align}
were $\rho$ is the macroscopic density and $\boldsymbol{u}$ the macroscopic velocity.

The time and space discretization 
yields an updating rule for the PDFs which can be split in a \emph{collision} and a \emph{stream step}
\begin{align}
	\tilde{f}_\alpha(\boldsymbol{x}, t + \Delta t) &= f_\alpha(\boldsymbol{x},t) - \omega [ f_\alpha(\boldsymbol{x},t)-f_\alpha^{eq}(\rho, \boldsymbol{u}) ]  \\
	f_\alpha(\boldsymbol{x}+\boldsymbol{c}_\alpha \Delta t, t+\Delta t) &= \tilde{f}_\alpha (\boldsymbol{x},t+\Delta t),
\end{align}
and is known to yield a second order accurate approximation of the
incompressible flow equations of a viscous fluid. The discrete scheme
is following an ideal gas equation of state where the pressure is given by
\begin{equation}
  p = \rho \cdot c_s^2,
  \label{eq:eqOfState}
\end{equation}
with the lattice \emph{speed of sound} $c_s$ as a model parameter (herein, $c_s = 1 / \sqrt 3 $).

\subsection{Free Surface Lattice Boltzmann Method (FSLBM)}
\label{sec:FSLBM}

\begin{figure}[h]
  \centering
  \includegraphics[width=0.4\textwidth]{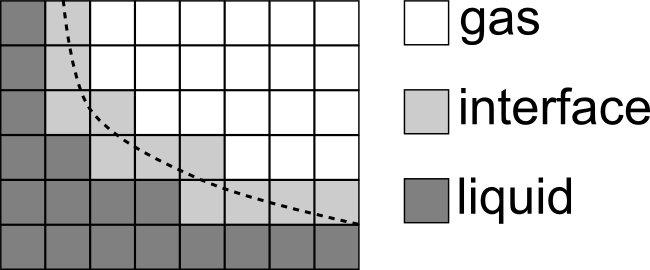}
  \caption{Different cell types. Only interface and liquid cells are
    included in the lattice Boltzmann scheme.}
  \label{fig:interface}
\end{figure}

The FSLBM used in the following is based on the assumption that the
simulated liquid-gas flow is governed by the first phase completely,
such that the dynamics of the gas phase can be neglected. Thus, the
problem is reduced to a single phase flow with a free boundary
\cite{KoernerEtAl}. The liquid phase is simulated with a single phase
LBM as described in Sec \ref{sec:LBM}. To treat the boundary at the
free surface between the two phases a volume of fluid approach is
used. The mass flux is tracked for so called \emph{interface} cells,
where a lattice cell is a cubic volume of unit length centered
around a lattice node. The interface cells form a closed layer between
the gas and the liquid cells, and hold a \emph{fill level} $\varphi$
which ranges in between zero and one, see Fig. \ref{fig:interface}.  A
fill level of $\varphi=0$ denotes an empty cell, $\varphi=1$
corresponds to a completely filled liquid cell. Thus, the liquid mass
$m$ can be computed for each cell with Equ. \ref{equ:mass}.
\begin{align}
\label{equ:mass}
 m = \varphi (\Delta x)^3 \rho
\end{align}
where $\rho$ is the local density.

For the interface cells the mass exchange is computed directly from
the stream step of the LBM, cf. Sec. \ref{sec:LBM}, and corresponds to a change
of the fill level. For an interface cell at $\boldsymbol{x}$ the mass balance with
a neighbor at $\boldsymbol{x}+\boldsymbol{c}_{\alpha}$ is given by
\begin{equation}
  \Delta m_{\alpha} =
  \begin{cases}
    0 & \text{if  $\boldsymbol{x}+\boldsymbol{c}_{\alpha}$ is gas,}\\
    f_{\bar{{\alpha}}}(\boldsymbol{x}+\boldsymbol{c}_{\alpha}, t) - f_{\alpha}(\boldsymbol{x}, t) & \text{if $\boldsymbol{x}+\boldsymbol{c}_{\alpha}$ is liquid,}\\
    \frac{1}{2} \left( \varphi(\boldsymbol{x}, t) + \varphi(\boldsymbol{x}+\boldsymbol{c}_{\alpha}, t) \right) \cdot \left( f_{\bar{{\alpha}}}(\boldsymbol{x}+ \boldsymbol{c}_{\alpha}, t) - f_{\alpha}(\boldsymbol{x}, t)  \right) & \text{if $\boldsymbol{x}+\boldsymbol{c}_{\alpha}$ is interface}.  
  \end{cases}
  \label{eq:massExchange}
\end{equation}

When the mass changes there is the possibility that the state of a
lattice cell changes between interface, fluid and gas cells if the
fill level approaches $1$ or $0$, respectively. This can be
facilitated such that the interface cells always form a closed layer
around the liquid phase (cf. Fig. \ref{fig:interface}), such that
appropriate boundary handling for the lattice Boltzmann scheme can be
done. Empty cells with $\varphi=0$ (gas) are excluded from the lattice
Boltzmann simulation.

\begin{figure}[h]
  \centering
  \includegraphics[width=0.2\textwidth]{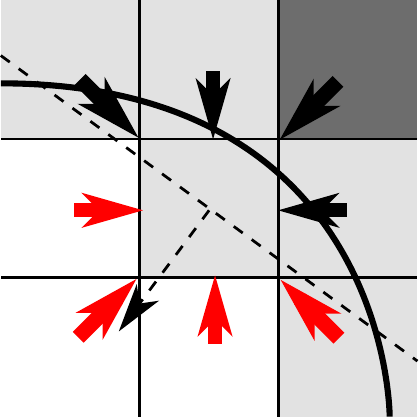}
  \caption{Missing particle distribution functions from gas phase.}
  \label{fig:reconstruct}
\end{figure}

To represent the behavior of a free surface, a special boundary
condition is applied.  Since the gas phase is not computed, there are
no PDFs available for the gas cells.  Hence, after the stream step,
the interface cells do not have a full set of PDFs. In Fig. \ref{fig:reconstruct} the
missing PDFs are marked in red.  The dashed line denotes a locally
reconstructed tangent to the free surface which is given by the bold
line.

The missing PDFs are reconstructed as suggested in \cite{KoernerEtAl}:
\begin{align}
  f_\alpha^{\text{liquid}}(\boldsymbol{x},t) &= f_\alpha^{\text{gas}}(\boldsymbol{x},t) + f_{\bar{\alpha}}^{\text{gas}}(\boldsymbol{x},t) - f_{\bar{\alpha}}^{\text{liquid}}(\boldsymbol{x},t).
\end{align}

Here the $f_\alpha^{\text{gas}}(\boldsymbol{x},t) =
f_\alpha^{\text{eq}}(\rho_{\text{gas}}(\boldsymbol{x}), \boldsymbol{u}(\boldsymbol{x}))$ are computed as
equilibrium with the current velocity at the boundary and the pressure
of the gas phase according to the LB equation of state,
Equ. \ref{eq:eqOfState}. This reconstruction is done for all directions
$\alpha$ that are oriented \emph{towards} a locally approximated
tangent plane of the free surface.
$\bar \alpha$ denotes the opposing direction to $\alpha$.

The gas pressure is composed of two parts, the bubble pressure $p_V$
(cf. Sec. \ref{sec:bubbleModel}) and the Laplace pressure $\Delta p_{\sigma}$.
\begin{align}
  p_{\text{gas}} = p_V + \Delta p_\sigma
  \label{eq:gasPressure}
\end{align}
The first part is given from the initial volume $V^*$ and the current volume $V(t)$ of the
bubble by
\begin{align}
  p_V = \frac{\text{initial bubble volume}}{\text{current bubble
      volume}}\cdot p_0 = \frac{V^{*}}{V(t)} \cdot p_0.
  \label{eq:idealGasLaw}
\end{align}
The initial pressure $p_0$ can be written as
\begin{align}
 p_0 = \rho_0 \cdot c_s^2
\end{align}
where $\rho_0$ is unity.

To get the information of the current bubble volume $V(t)$, a detailed
tracking of the volume of each bubble is done by the bubble model as
described in Sec. \ref{sec:bubbleModel}.
The second term of Equ. \ref{eq:gasPressure} is necessary to capture
surface effects due to surface tension.  The Laplace pressure is
computed as
\begin{align}
  \Delta p_\sigma = 2 \sigma \kappa(\boldsymbol{x},t),
\end{align}
where $\sigma$ is the surface tension constant of the interface and
$\kappa(\boldsymbol{x},t)$ is the local curvature. We approximate $\kappa$ by a
local triangulation of the interface based on \cite{ParkerYoungs},
with details described in \cite{Pohl2007}.

\subsection{Bubble Model}
\label{sec:bubbleModel}
In order to track the unconnected gas regions as distinct bubbles with
an individual gas pressure, a bubble model is needed. As described in
\cite{Pohl2007,Donath2009}, every bubble is
represented by a data set which holds information like the current
and initial volume and is identified by a unique bubble ID. The
current volume of a bubble is updated in every time step from the
changes of the fill levels of its surrounding interface cells
according to Equ. \ref{eq:massExchange}. From that the bubble pressure
is updated and taken into account in the free surface boundary
condition of the FSLBM, cf. Sec. \ref{sec:FSLBM}.

Additional complexity arises, whenever the connectivity of the gas
subdomain changes, i.e., a coalescence or breakup of a bubble needs to
be treated. In order to detect all gas regions within the fluid,
\cite{Cobussat2005} applies an implicit solver on the fill level
information, and therefrom updates the connectivity information of
bubbles as well as each bubble's individual volume and pressure. This
is done at each time step and somewhat costly, but reportedly cheaper
than solving compressible Euler equations for the gas phase. The
bubble model presented by us is different in the way of how connected
gas regions are tracked: Instead of implicitly detecting and numbering
the bubbles, a specially adapted \emph{flood fill} algorithm
\cite{FloodFill} is used to extract the connectivity information and
update the bubble data. Since, the pressure changes in the bubbles due
to interface advection can be computed directly from the fill level
balance, Equ. \ref{eq:massExchange}, in each time step, the flood fill
algorithm is needed only if the connectivity of the gas-subdomain
changes. I.e., for the vast majority of computed time steps it can be
skipped completely and thus has no significant impact on the overall
performance of the simulation. The enhanced bubble model presented in
the following distinguishes the case of a \emph{merge} (coalescence)
of multiple bubbles into one bubble (Sec. \ref{sec:merge}) from
the case of a \emph{split} of a single bubble into several daughter
bubbles (Sec. \ref{sec:split}).

\subsubsection{Bubble Merge}
\label{sec:merge}
The case of bubble coalescence has already been treated in
\cite{Pohl2007,Donath2009}. If $B$ is the set of coalescing
bubbles, with initial volumes $V^*_b$ ($b \in B$) and current volumes
$V_b(t)$ ($b \in B$), a new bubble $b_{\text{new}}$ is created, to
replace the bubbles in $B$. The new bubble is initialized with initial
volume $V^*_{ b_{\text{new}} } = \sum_{ b \in B } V^*_b $ and current
volume $V_{ b_{\text{new}} }(t) = \sum_{ b \in B } V_b(t), $ and gas
pressure according to Equ. \ref{eq:idealGasLaw}.  Algorithmically,
merges are detected by comparing the bubble IDs of neighboring
interface cells. If the next neighbor of an interface cell is itself
an interface cell belonging to a separate bubble, a merge is
triggered. The breakup of a bubble into several pieces is more
complicated, as described in the next section.

\subsubsection{Bubble Split}

\begin{figure}[h]
  \subfloat[Anti-parallel normals \label{fig:normals}]{\includegraphics[width=.28\textwidth]{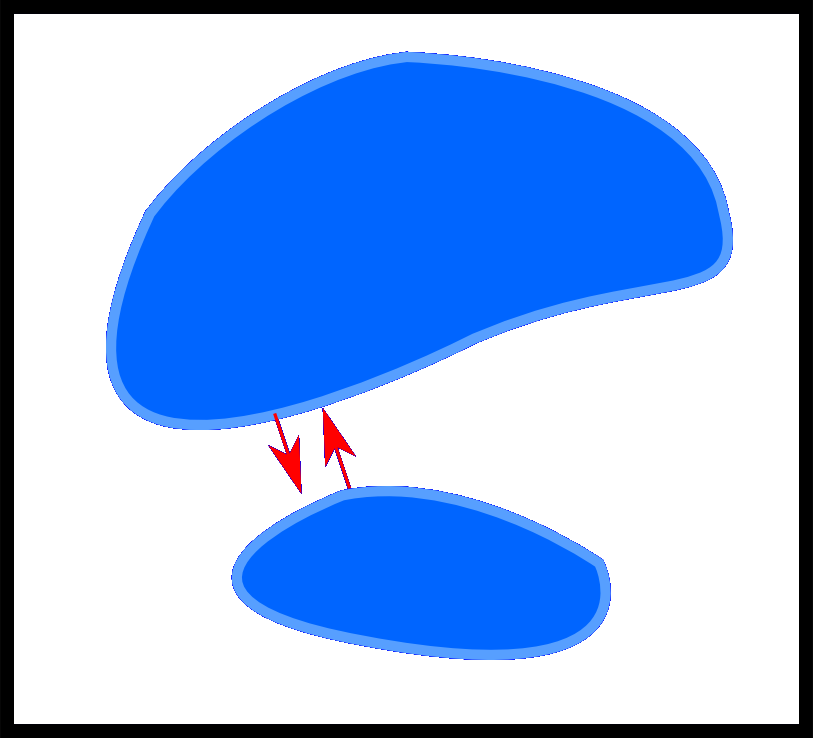}}\hfill
  \subfloat[Seed-fill \label{fig:seedfill}]{\includegraphics[width=.28\textwidth]{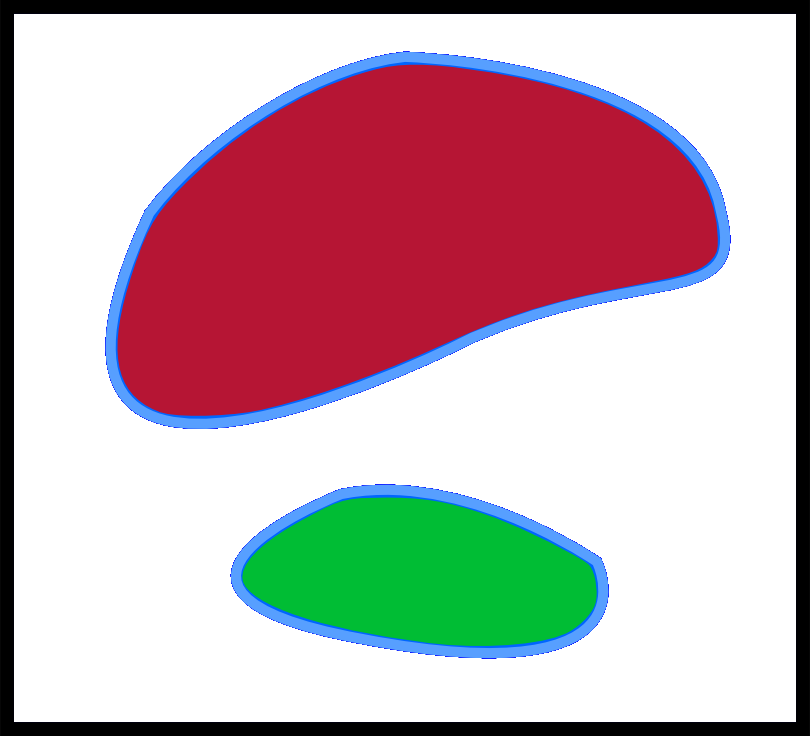}}\hfill
  \subfloat[New volumes \label{fig:newVols}]{\includegraphics[width=.28\textwidth]{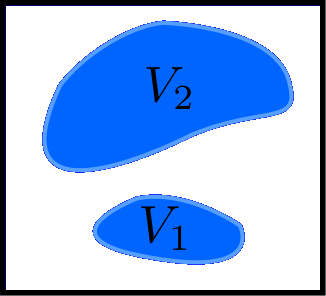}}
  \caption{Serial bubble splitting: Anti-parallel surface normals in the neighborhood of a liquid film (a) trigger the bubble split detection of Sec. \ref{sec:split} which colors the connected sub-volumes of the bubble (b). If more then one regions are found, the bubble has split, and each sub-volume defines a new bubble (c).}
  \label{fig:serialSplit}
\end{figure}

\label{sec:split}
Since the pressure of each bubble is updated according to the ideal
gas law, Equ. \ref{eq:idealGasLaw}, in case of a breakup of a bubble,
the sub-volume of each daughter bubble needs to be determined. For a
bubble $b_{ \text{old} }$ that breaks up into a set $B$ of daughter
bubbles with volumes $V_b(t)$ ($b \in B$), new bubble data entries
are created: The current pressure for each daughter bubble is set to
$p_b(t) = p_{ \text{old} }(t)$ ($b \in B$), and the initial volumes are set to
$V^*_b = V_b(t)$ ($b \in B$). The information of the former bubble $b_{
  \text{old} }$ is no longer needed for succeeding time steps.
However, the determination of the sub-volumes $V_b(t)$ ($b \in B$) is
non-trivial, and needs to be extracted from the fill level information
in the grid. This is achieved by a special \emph{bubble split
  detection} that consists of an iterative flood fill algorithm, which
runs over the grid and colors the connected subregions of the former
bubble $b_{ \text{old} }$ with distinct colors. Every color then
represents an unconnected gas region as a result of the flood fill
algorithm \cite{FloodFill}. Thereby, also the volume of each subregion
is computed by accumulating the gas cells and the interface cells
according to their fill levels. Thus, after the termination of the
algorithm, the number of daughter bubbles including all information
(bubble volume and pressure) needed for further processing is
known. 

As indicated in Fig. \ref{fig:normals}, it is sufficient to trigger
the bubble split detection only if a grid configuration occurs, where
in the neighborhood of a liquid cell, two interface sections of the
same bubble have anti-parallel surface normals. Depending on the
simulated problem, the split algorithm is triggered only for a small
number of time steps. This split detection has first been implemented
by \cite{Siko} for serial computations. Implementation details for
parallel computing are given below.

\subsection{Implementation Details}
\label{sec:impl}
In order to identify the disconnected gas areas, namely the bubbles,
they are identified by bubble IDs. For parallel computations the
domain is split up spatially into several blocks to be distributed to
a number of processing units. This mapping is usually one-to-one. In
addition, every processor holds the bubble data of all bubbles
residing on its blocks. Note, that it is well possible for a single
bubble to extend over multiple blocks and processors
\cite{Donath2009}. Therefore, the information update for a bubble may
degenerate into a concurrent task. In order to update the bubble data
for each bubble coherently on all the involved processors, the bubble
data entry for each bubble also holds a list of all blocks it is
overlapping with. If a bubble is vanishing from a block or advected to
another, this data has to be adapted carefully.

For example, Fig. \ref{fig:secondComm} shows two bubbles denoted in green and violet, respectively. 
Process number one and two only hold the data for the green bubble. 
Process three and four hold the bubble data sets for the green and the violet bubble.
The green bubble holds the information that it resides on processes one, two, three and four and the violet bubble processes three and four respectively.

In addition to the PDF data for the lattice Boltzmann scheme and the fill level information for the free surface algorithm, bubble IDs are stored in a field for all gas and interface cells \cite{Donath2009}.
Hence for each interface or gas cell the corresponding bubble data can directly be accessed.
All blocks are enclosed by an additional layer of cells which is denoted \emph{ghost layer} or \emph{halo layer} \cite{Donath2009,KoernerEtAl2006}.
This additional cells are needed for communication in general and herein specifically for the communication of color information in the parallel breakup algorithm.
In case of a split, those fields have to be updated with the newly generated bubble IDs.
Similar fields are also needed to store the colors of the flood fill
algorithm described in Sec. \ref{sec:split}. The colors for the flood fill algorithm are realized as bubble IDs.




\subsubsection{Parallel Algorithm}
\label{sec:parAlgo}

\begin{figure}[h]
 \subfloat[Detect breakup \label{fig:detBu}]{\includegraphics[width=.23\textwidth]{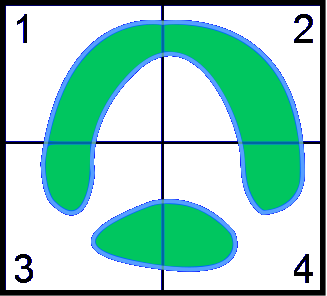}}\hfill
 \subfloat[Local seed-fill \label{fig:locSeedfill}]{\includegraphics[width=.23\textwidth]{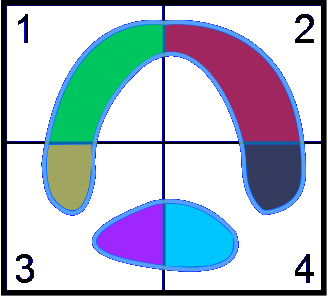}}\hfill
 \subfloat[First communication \label{fig:firstComm}]{\includegraphics[width=.23\textwidth]{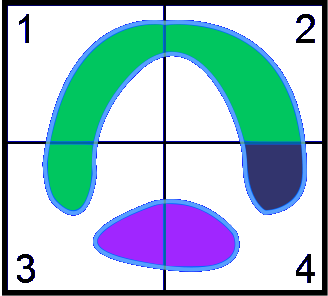}}\hfill
 \subfloat[Second communication \label{fig:secondComm}]{\includegraphics[width=.23\textwidth]{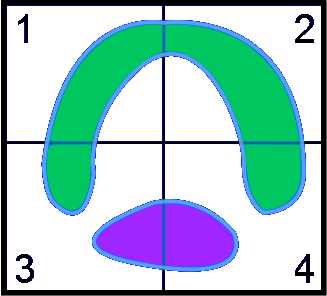}}
 \caption{Parallel bubble splitting}
\end{figure}

In the parallel case the algorithm has to account for bubbles residing
on more than one process and communicate over process borders as explained above.  
In Fig. \ref{fig:detBu}, a bubble residing on four processors has split
twice and two daughter bubbles have developed. For the split recognition,
every process does a local flood fill algorithm as described in
Sec. \ref{sec:split}. This results in a possibly inconsistent
coloring of the bubble (cf. Fig. \ref{fig:locSeedfill}) where one potentially connected gas region is colored differently.
The upper gas region residing on four processes is distributed over four color regions and the lower one over two.
So, consequently, the color information has to be unified by communication
between the neighboring processes. The ghost layers are subsequently sent to the neighboring processes which makes the information about the colors from the adjacent process available there, and mismatching colors within connected regions can be detected.
After the first communication the situation is visualized in Fig. \ref{fig:firstComm}.  
The processes which are adjacent to each other decide on one color and update their color information respectively.
The process with the lowest number goes first.  
In the next iteration, the color information from the next neighbors are reached and again the color is adapted
accordingly, see Fig. \ref{fig:secondComm}.
The iteration terminates, if no more color mismatches are found at the process borders.
The maximal number for those communication steps is thus limited by the number of processes that the gas region
resides on. 

After all the communication is done, new data sets for the daughter bubbles can be created according to the remaining colors. 

\section{Examples and Validation}
\label{sec:exp}
\subsection{Rupture of a thin bubble}

\begin{figure}[h]
  \framebox[\linewidth]{ \includegraphics[scale=0.27]{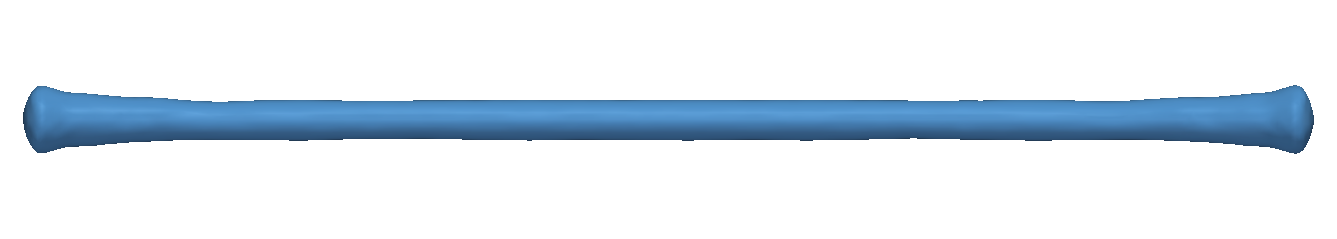} $t=22 ms$ }
  \framebox[\linewidth]{ \includegraphics[scale=0.27]{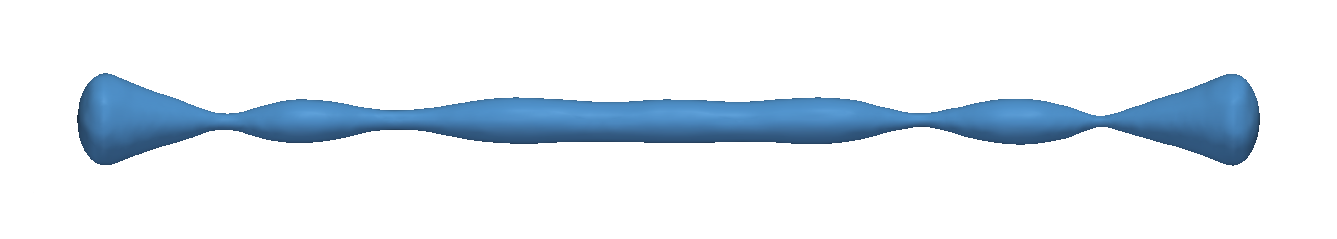} $t=38 ms$ }
  \framebox[\linewidth]{ \includegraphics[scale=0.27]{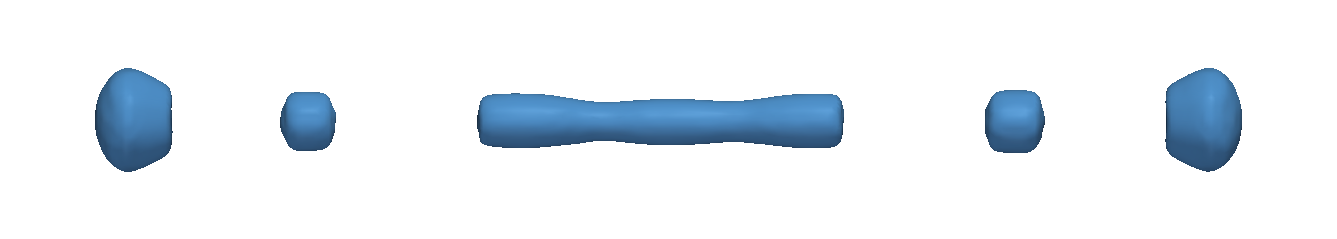} $t=46 ms$ }
  \framebox[\linewidth]{ \includegraphics[scale=0.27]{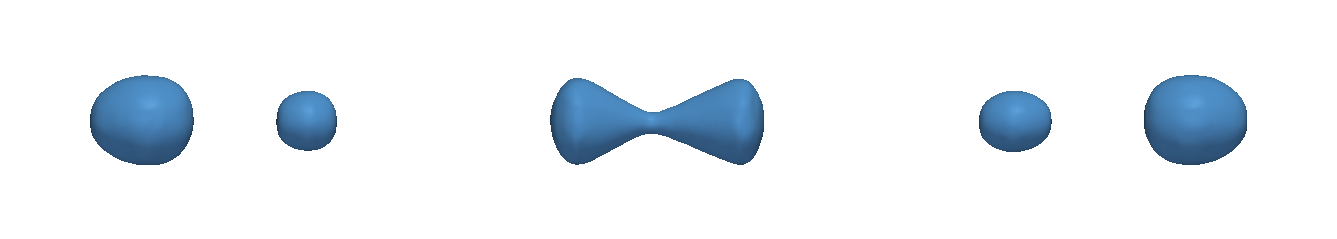} $t=68 ms$ }
  \framebox[\linewidth]{ \includegraphics[scale=0.27]{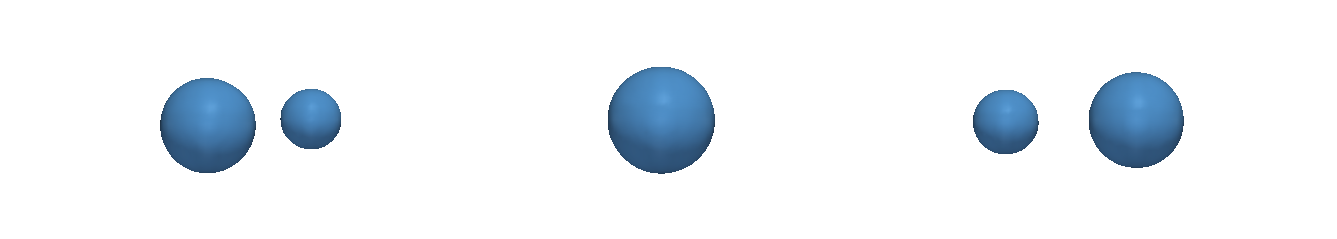} $t=324 ms$ }
  \caption{Rupture of a thin bubble at selected time steps. Surface tension is $\sigma=72mN/m$. In this case, the bubble breaks into five pieces.}
  \label{fig:rup}
\end{figure}

To demonstrate the functionality of the enhanced bubble model, we simulate the breakup of an initially threadlike bubble due to capillary forces.
A domain of $360 \times 40 \times 40$ lattice cells is initialized with a rectangular region of gas of size $336 \times 8 \times 8$ lattice cells.
The physical space step is $\Delta x = 0.125\times 10^{-4} m$ which results in a gas thread thickness of roughly $0.1 mm$.
In the two simulated cases, the surface tension is varied significantly.
The first simulation, see Fig. \ref{fig:rup}, is done for the physical quantities of an air water system.
The kinematic viscosity of the liquid is chosen $\nu = 10^{-6} m^2/s$ and the density $\rho = 1000 kg/m^3$. 
The surface tension is, as for an water-air system, $\sigma = 72 mN/m$.
Several spherical shaped bubbles are developing and the separation of those gas regions is correctly handled due to the splitting algorithm which initializes the new volumes correctly after every split.
The process of breaking up is taking place within milliseconds.

\begin{figure}[h]
  \framebox[\linewidth]{ \includegraphics[scale=0.27]{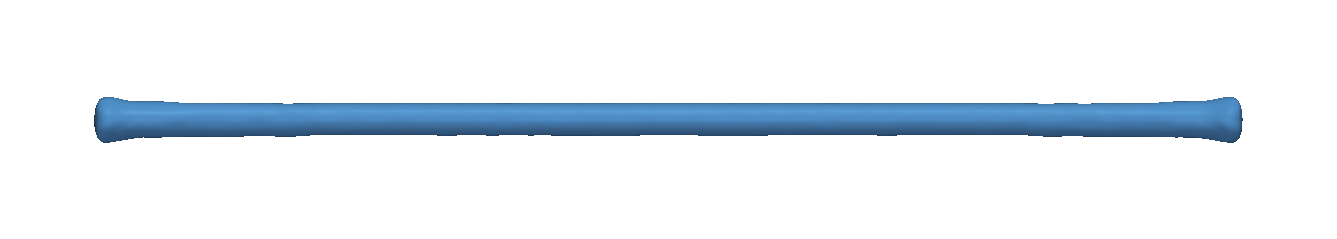} $t=22 ms$ }
  \framebox[\linewidth]{ \includegraphics[scale=0.27]{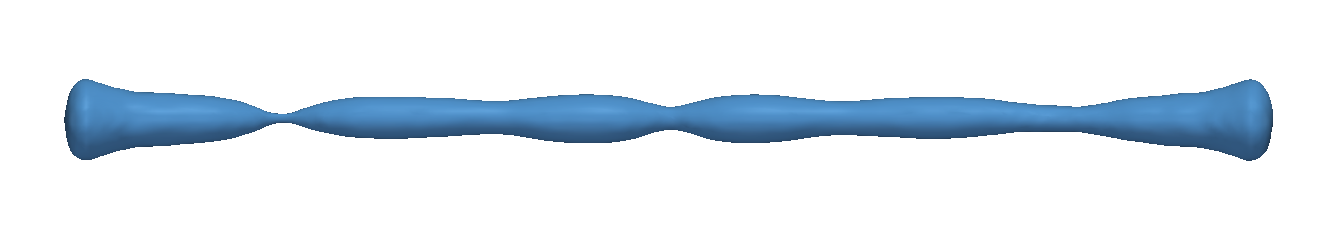} $t=65 ms$ }
  \framebox[\linewidth]{ \includegraphics[scale=0.27]{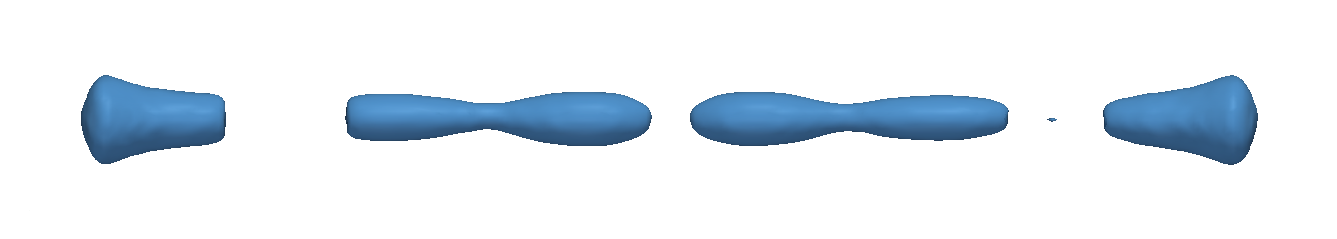} $t=76 ms$ }
  \framebox[\linewidth]{ \includegraphics[scale=0.27]{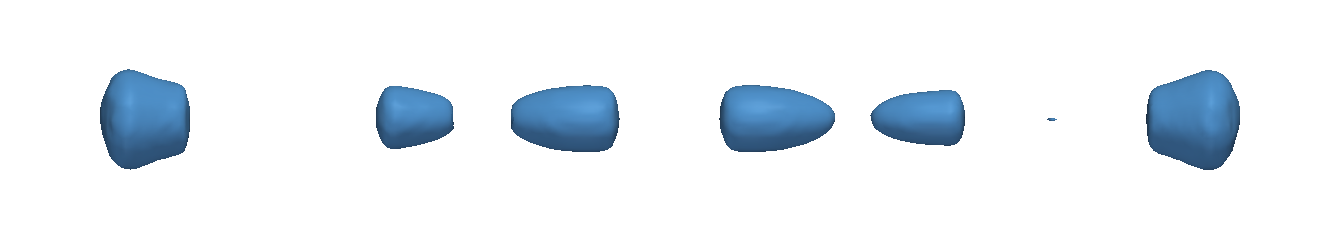} $t=90 ms$ }
  \framebox[\linewidth]{ \includegraphics[scale=0.27]{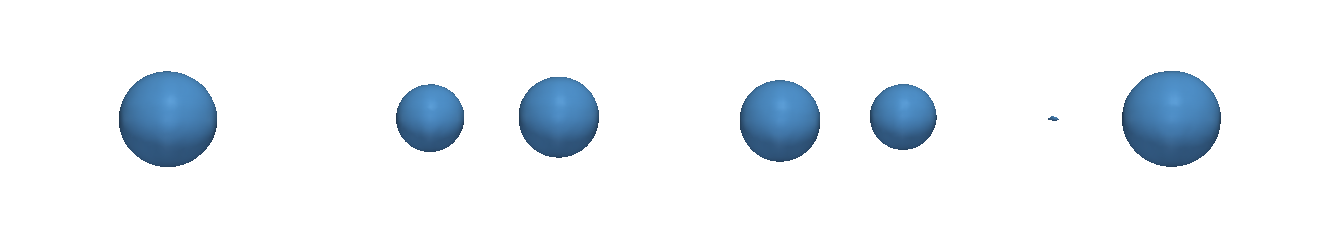} $t=324 ms$ }
  \caption{Rupture of a thin bubble at selected time steps. Surface tension is $\sigma=18mN/m$. In this case, the bubble breaks into six pieces.}
  \label{fig:rup2}
\end{figure}

The second simulation runs with a reduced surface tension parameter of $\sigma=18mN/m$ and can be seen in Fig. \ref{fig:rup2}.
This is four times smaller than in the previous simulation. 
Again the surface tension effects forces the threadlike bubble to break into several bubbles.

This shows that the enhanced bubble model is capable of treating the
breakup of bubbles correctly.
Similar simulations with a droplet splitting in a shear flow have been done in \cite{Renardy2000}.

\subsection{Bubble detachment from a pore}

\begin{figure}[h]
 \subfloat[$t=2.2ms$ \label{fig:det1}]{\framebox[.24\textwidth]{ \includegraphics[width=.24\textwidth]{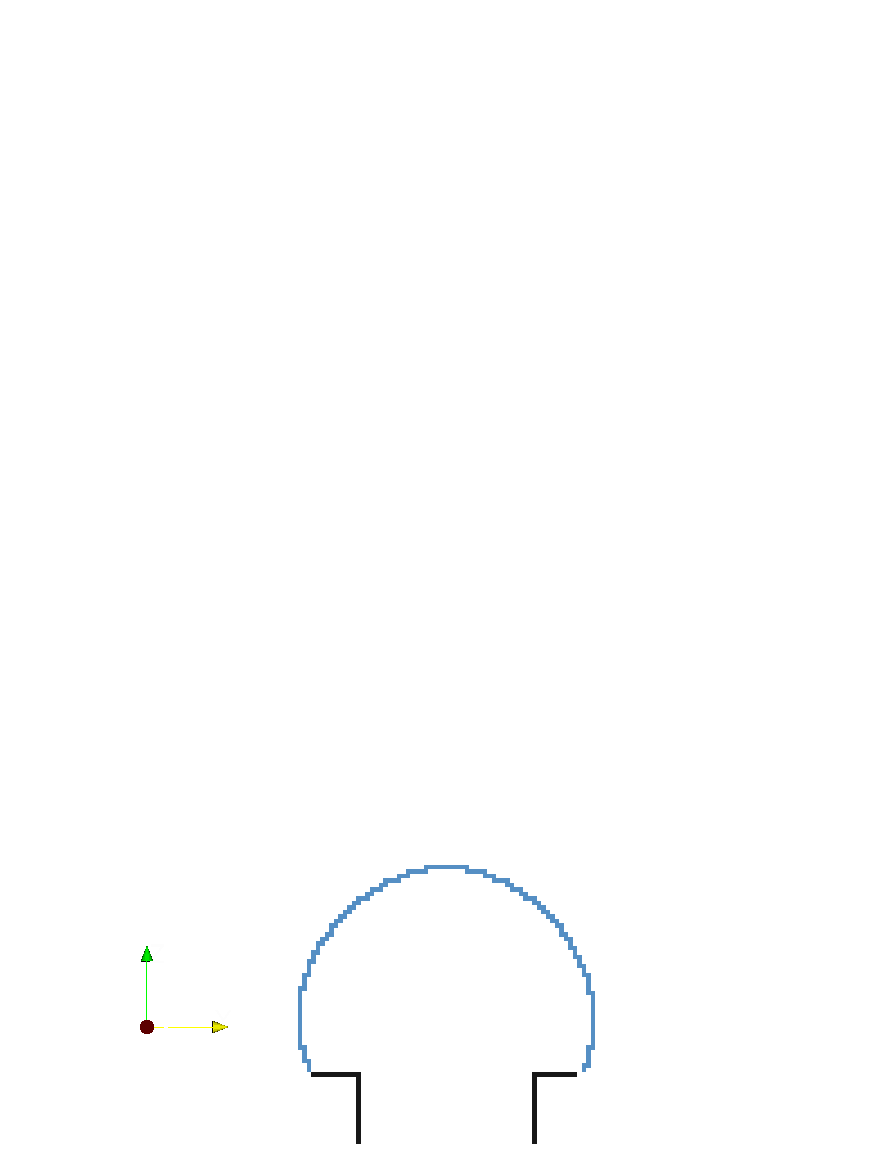}}}\hfill
 \subfloat[$t=54ms$ \label{fig:det2}]{\framebox[.24\textwidth]{ \includegraphics[width=.24\textwidth]{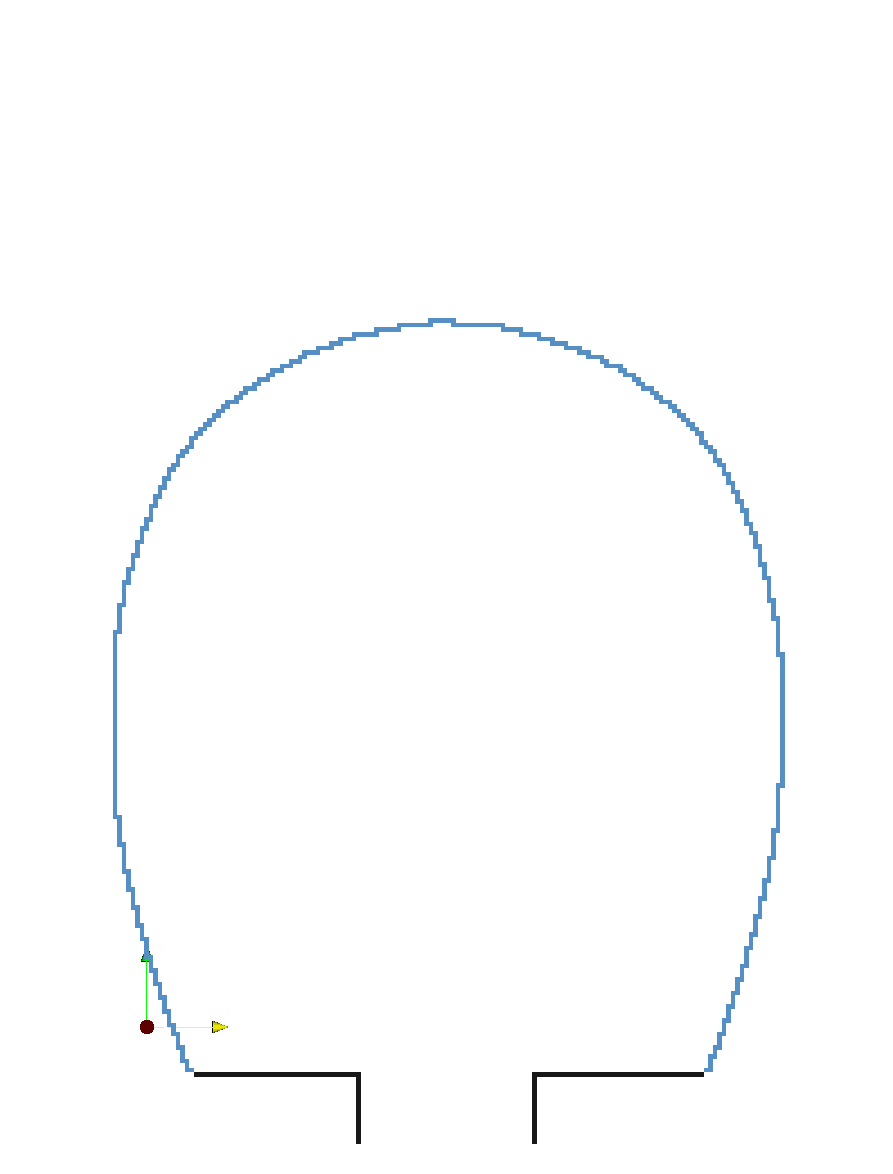}}}\hfill
 \subfloat[$t=76ms$ \label{fig:det3}]{\framebox[.24\textwidth]{ \includegraphics[width=.24\textwidth]{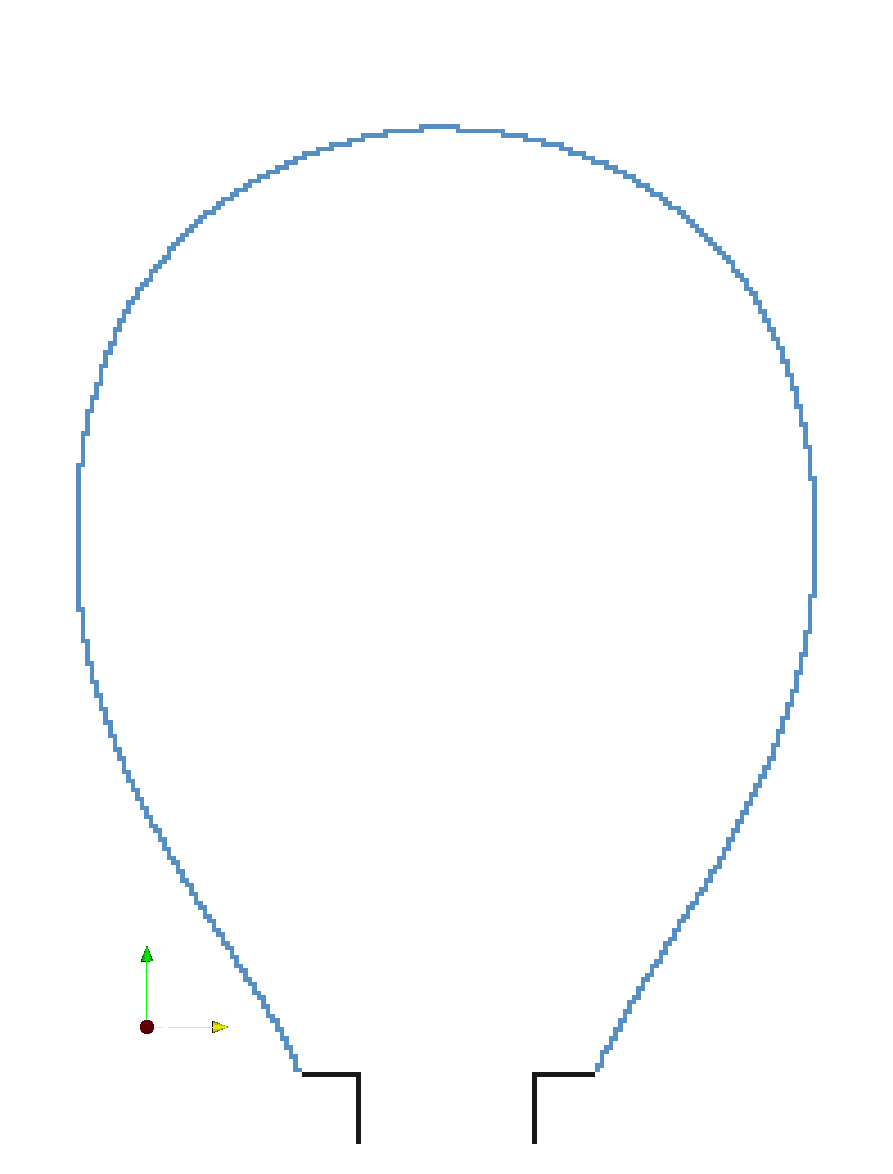}}}\hfill
 \subfloat[$t=87ms$ \label{fig:det4}]{\framebox[.24\textwidth]{ \includegraphics[width=.24\textwidth]{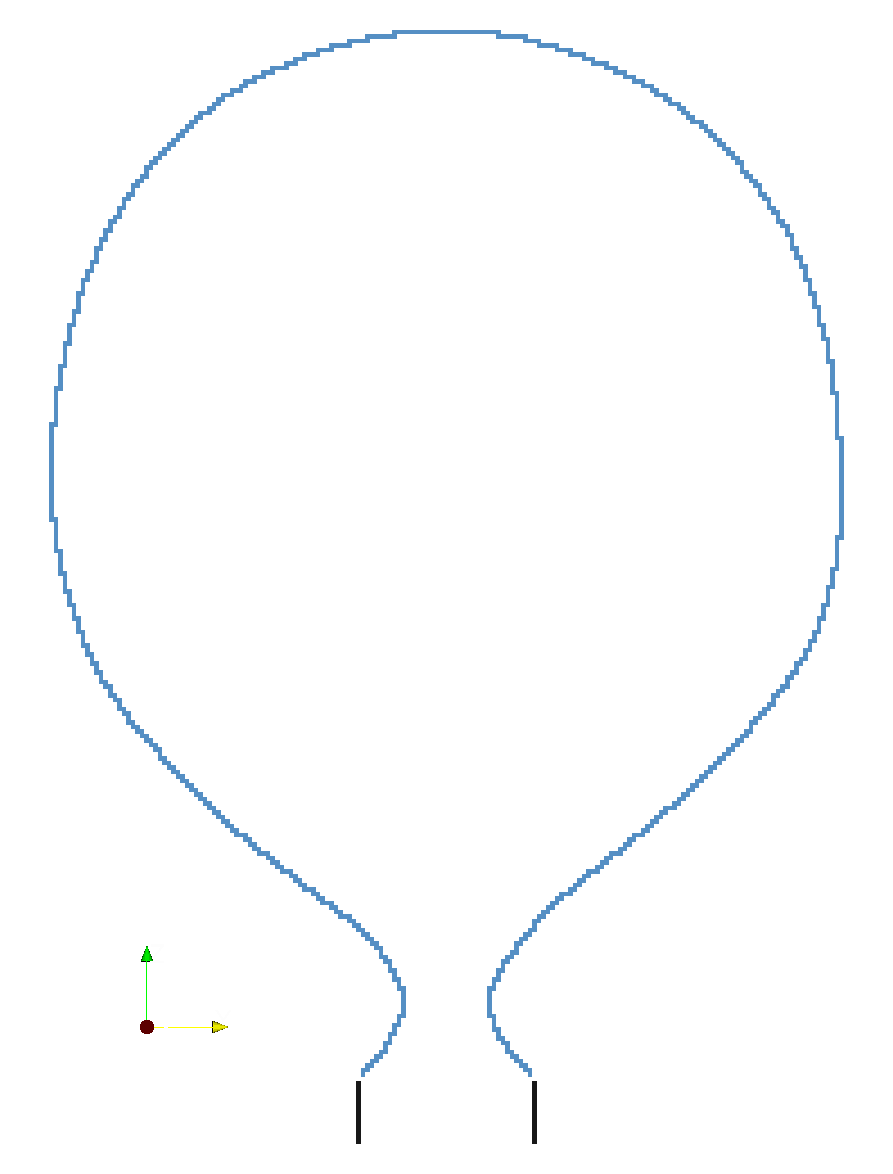}}}
\caption{Bubble development at an orifice over time with $\theta = 70^\circ$ and inflow rate $10ml/min$. A cut is made through the $yz$-plane in the center. The blue line denotes the interface cells and the black line the solid border.}
\end{figure}

A further application of the advanced bubble model is the detachment
of bubbles from a pore.  The volume of the detaching bubble
was investigated in simulations. In \cite{Gerlach2005} the bubble formation of a
submerged orifice at low gas inflow rates ($\dot{Q} \rightarrow 0$) is
investigated analytically. Through the submerged pore in a quiescent
liquid, gas is injected at a constant rate. In the \emph{quasi-static}
regime, the developing gas bubble detaches from the pore, as soon as
the buoyancy forces outbalance the capillary forces.
\cite{Gerlach2007} addresses the problem numerically with a
two-dimensional combined level set - volume of fluids code. With the
FSLBM, similar simulations were carried out for a water-air system
with the density of the liquid $\rho=998.12kg/m^3$, the dynamic
viscosity of the liquid $\eta = 1.002\times 10^{-3}Pa\, s$ and the
gravitational acceleration $g=9.81m/s^2$. The radius of the orifice is
$r=1mm$ and the domain size $228 \times 228 \times 450$. Non-slip
boundary conditions were used at bottom and side walls of the
domain. An open boundary condition is applied at the top.  The
physical space step is $\Delta x = 0.\overline{3}\times 10^{-4}m$ and the
resolution of the orifice is 30 lattice cells. Changing the resolution of the orifice to 20 and 40 lattice cells, 
respectively, while keeping the physical radius by changing the space step, yields deviations in the detachment volume below $2.6 \%$.
A certain discretization error is expected in the 3D FSLBM simulation,
which uses a Cartesian grid to approximate the circular pore, whereas
the 2D simulations in \cite{Gerlach2007} exploit the radial symmetry of
the problem. 
The static contact angle $\theta$ which is a property of the orifice material was varied in order to compare the data
with the literature. $\theta$ is the angle between solid and interface in the liquid phase which is reached by a three phase system in equilibrium. 
During the development of the bubble, the system runs through various dynamic contact angles always tending to capture $\theta$ which enforces the contact line to also leave the border of the orifice.
The numerical contact angle model used for the simulation is explained in \cite{DonathDiss}.
In Fig. \ref{fig:det1}-\ref{fig:det4} the
development over time for $\theta = 70^\circ$ and an
inflow rate of the air of $10ml/min$ is shown.
Tab. \ref{tab:detRad} shows the detachment volumes $V_d$ in comparison to
the available literature. The results obtained with the FSLBM with
enhanced bubble model are in good agreement at both material
parameters. An increase of the gas inflow rate was found to result in
increased detached bubble volumes as reported in \cite{Gerlach2007}.

\begin{table}[h]
\centering
\begin{tabular}{lccc}
\toprule
$\theta$ & \multicolumn{3}{c}{$V_d$} \\
\cmidrule(r){2-4}
& FSLBM & 2D VOF \cite{Gerlach2007} & Analytic \cite{Gerlach2005} \\
\cmidrule(r){2-4}
 & $1ml/min$  & $1ml/min$ & $\dot{Q} \rightarrow 0$ \\
\midrule
$70^\circ$ & $38 mm^3$  & $39 mm^3$ & $33 mm^3$\\
$90^\circ$ & $51 mm^3$ & $69 mm^3$ & $69 mm^3$\\
\bottomrule
\end{tabular}
\caption{Results for the detachment volume $V_d$ for different static contact angles $\theta$.}
\label{tab:detRad}
\end{table}

\section{Conclusion}
\label{sec:conclusion}

This paper presents the algorithm for an enhanced bubble model that extends an existing single phase free surface approach to support the simulation of bubbly flows.
The bubble model simulates the gas pressure according to ideal gas law but neglects the gas dynamics which saves computational costs.
The underlying numerical method for the hydrodynamics is the lattice Boltzmann method with a volume of fluids approach for interface capturing.
By introducing a new algorithm, it is possible to detect topological changes of the free surface and thus register the breakup as well as the coalescence of bubbles.
To support large scale simulations, we also present details of a parallel implementation.

To show the mode of action of the breakup, two test cases are presented.
The first setup shows the rupture into multiple bubbles from a threadlike bubble for different surface tension values.
The second scenario is a bubble detaching from a pore.
Here the simulations are done with changing inflow rates and static contact angles. 
The detached bubble volumes are compared to those of \cite{Gerlach2007} and reveal good agreement.

In \cite{KoernerEtAl} foam development has already been investigated with an equivalent free surface lattice Boltzmann approach and a simplified bubble model.
With the enhanced bubble model, including the correct treatment of bubble breakup, also generation of foam, e.g. at a membrane, could be simulated.
This will be subject of future work.

\section{Acknowledgements}
For the simulations presented in this paper, the \emph{waLBerla}
lattice Boltzmann framework \cite{walberla} has been used. The authors
would like to thank all members of the waLBerla team.
This research project was supported by the German Ministry of Economics and Technology (via AiF) and the FEI (Forschungskreis der Ernährungsindustrie e.V., Bonn). Project AiF 17125 N.

\bibliographystyle{plainnat}
\bibliography{lit}

\end{document}